\documentclass[12pt]{article}
\usepackage{amsfonts}
\usepackage{amssymb}
\def\a{\alpha}
\def\b{\beta}
\def\g{\gamma}
\def\d{\delta}

\def\l{\lambda}
\def\o{\omega}
\def\p{\partial}

\def\vn{\varnothing}
\def\vp{\varphi}
\def\s{\sigma}
\def\th{\theta}

\def\su{\subset}

\newcommand{\ce}{{\mathcal E}}

\newcommand{\cg}{{\mathcal G}}
\newcommand{\ch}{{\mathcal H}}

\newcommand{\co}{{\mathcal O}}

\newcommand{\cs}{{\mathcal S}}
\newcommand{\mr}{{\mathcal R}}

\newcommand{\cv}{{\mathcal V}}
\newcommand{\U}{{\mathcal U}}
\newcommand{\C}{\mathbb C}
\newcommand{\R}{\mathbb R}
\newcommand{\Z}{\mathbb Z}
\newcommand{\im}{{\mathop{\mbox{Im}}\nolimits\,}}
\newcommand{\Ker}{{\mathop{\mbox{Ker}}\nolimits\,}}
\newcommand{\Tr}{{\mathop{\mbox{Tr}}\nolimits\,}}

\begin{document}
\setlength{\baselineskip}{5mm}

\begin{flushright}
 hep-th/0101150
\end{flushright}

\begin{center}
\noindent{\large\bf
\v{C}ech, Dolbeault and de~Rham cohomologies\\
\vspace{2mm}
in Chern-Simons and BF theories
\footnote{To appear in proceedings of the XXIII International
Colloquium on Group Theoretical Methods in Physics
(July 31 - August 5, 2000, Dubna, Russia).}
}

\vspace{4mm}

\noindent{
T.A.Ivanova{\small$^a$}
and {
{A.D.Popov}}{\small$^{a,b}$}
}\vspace{1mm}

\noindent{\small
$^a$Bogoliubov Laboratory of Theoretical Physics, JINR, Dubna, Russia \\
$^b$Institut f\"ur Theoretische Physik, Universit\"at Hannover, Germany
}
\end{center}

\begin{quote}

{\small
{\sf Abstract.}
Topological Chern-Simons (CS) and BF theories and their
holomorphic analogues are discussed in terms of de Rham
and Dolbeault cohomologies. We show that \v{C}ech cohomology
provides another useful description of the above topological
and holomorphic field theories. In particular, all hidden
(nonlocal) symmetries of non-Abelian CS and BF theories can be
most clearly seen in the  \v{C}ech approach. We consider
multidimensional Manin-Ward integrable systems and describe
their connections with holomorphic BF theories. Dressing
symmetries of these generic integrable systems are briefly
discussed.
}

\end{quote}

{\bf 1. Introduction}

\smallskip

Topological field theories [1] were intensively studied
over the last ten years (see e.g. [2,3] and references
therein). Among these theories are Chern-Simons [4,5]
and BF [4,6,7] topological theories describing flat
connections $d_A= d+A$ on principal bundles $P$ over
smooth manifolds and $d_A$-closed ad$P$-valued forms.
Holomorphic Chern-Simons [8-10] and holomorphic BF [11]
theories describe flat (0,1)-connections $\bar\p_A=\bar\p
+ A^{0,1}$ on bundles $P$ over complex manifolds
and $\bar\p_A$-closed ad$P$-valued forms.
In Abelian case the above theories give field-theoretic descriptions
of de Rham and Dolbeault cohomologies.

After recalling the definitions of de Rham, Dolbeault and \v{C}ech
complexes [12,13], we discuss isomorphisms between the de Rham and \v{C}ech
cohomologies of real manifolds and the Dolbeault and \v{C}ech
cohomologies of complex manifolds. These isomorphisms permit one
to reduce differential equations of motion of CS and BF theories
defined on a manifold $M$ to some functional equations defined on
open subsets of $M$. Transition from the description of CS and BF
theories in terms of globally defined differential forms  (de Rham
or Dolbeault cocycles) to the description in terms of locally defined
functions (\v{C}ech cocycles) is especially useful in considering
moduli spaces of locally constant and holomorphic structures
(flat  connections and (0,1)-connections, respectively) on
bundles [14,15,10,16,17]. The \v{C}ech approach considerably simplifies
finding and exploring symmetries of Chern-Simons, BF and self-dual
Yang-Mills (SDYM) theories [16-18].

We describe a connection between holomorphic BF theories and
multidimensional integrable systems introduced by Yu.Manin [19]
and R.Ward [20] (see [21-23] for further developments).
These generic integrable systems contain 4D SDYM
model and 2D integrable models as special cases [19,20,23,24].
The equivalence of \v{C}ech and Dolbeault descriptions of holomorphic
bundles permits one to formulate a method of solving the Manin-Ward
equations as a cohomological analogue of the method of Birkhoff
factorizations and dressing transformations. Moreover,
by considering these generic integrable systems, it may be seen how the
dressing approach to integrable systems originates from deformation
theory of flat and holomorphic bundles. Dressing symmetries of
Manin-Ward integrable systems can be obtained from symmetries of CS and BF theories
described in [16,17].

\bigskip

{\bf 2. Differential complexes}

\smallskip

In this section we recall some definitions [12,13] to be used
in the following and fix the notation.

\smallskip

{\sf 2.1. General definitions.} A direct sum
$
C^* =\mathop{\oplus}\limits_{k\in\Z} C^k
$
of vector spaces $C^k$ indexed by integers $k$ is called a
{\it differential complex} if there are homomorphisms
$$...\to C^{k-1}\stackrel{\d}{\to}C^k\stackrel
{\d}{\to}C^{k+1}\to ...$$
such that $\d^2=0$. The homomorphism $\d$ is called a {\it differential
operator}  of the complex $C^*$. Elements $c^k\in C^k$ are called
{\it $k$-cochains}.

Let us consider the space
$$
Z^k:=\Ker\d\cap C^k =\{z^k\in C^k\ \mbox{such that}\ \d z^k=0\}\ .
\eqno(1)
$$
Elements $z^k\in Z^k$ are called {\it $k$-cocycles} and $Z^k$ is called
the {\it space of $k$-cocycles}. The space
$$
B^k:=\im\d\cap C^k =\{b^k\in C^k\ \mbox{such that}\ b^k=\d c^{k-1}\
\mbox{for some}\ c^{k-1}\in C^{k-1} \}
\eqno(2)
$$
is called the {\it space of $k$-coboundaries} and
elements $b^k\in B^k$ are called {\it $k$-coboundaries}.
It is clear that
$
B^k\su Z^k\su C^k\ ,
$
since each $k$-coboundary $b^k$ is a $k$-cocycle: $\d b^k=\d^2c^{k-1}=0$.
Cocycles $z^k$
such that $z^k\ne\d c^{k-1}$ are called {\it nontrivial} $k$-cocycles.
The space of nontrivial $k$-cocycles is parametrized by the quotient
space
$$H^k:=Z^k/B^k\ ,
\eqno(3)
$$
which is called the {\it $k$-th cohomology space} of the complex $C^*$.

The cohomology of the differential complex $C^*$
is the direct sum of the quotient spaces $H^k$,
$$H^*=\mathop{\oplus}\limits_{k\in\Z}H^k\ .
\eqno(4)
$$
De~Rham, Dolbeault and \v{C}ech complexes are examples
of the differential complex.

\medskip

{\sf 2.2. De~Rham complex.}
We consider a differentiable (smooth) manifold $M$ of real dimension $n$,
the space $\Omega^k(M)$ of smooth $k$-forms on $M$ and
an exterior derivative $d: \Omega^k(M) \to \Omega^{k+1}(M)$,
$d^2=0$. All that gives us the {\it de~Rham complex}
$$\Omega^*(M)=  \mathop{\oplus}\limits_{k=0}^n \Omega^k(M)$$
on a smooth $n$-manifold $M$.

A differential form $\omega$ on $M$ is called {\it closed} if $d\omega =0$.
A differential form $\tau$ on $M$ is called {\it exact } if $\tau =d\vp$ for some form
$\vp$.
Denote by $Z^k_d(M)$ the space of closed $k$-forms on $M$ and by $B^k_d(M)$
the space of exact $k$-forms on $M$, $B^k_d(M)\su Z^k_d(M)\su \Omega^k(M)$.
In the language of differential complexes, closed and exact $k$-forms
are called {\it de~Rham $k$-cocycles} and {\it $k$-coboundaries}, respectively.

The quotient space $H^k_d(M) = Z^k_d(M)/B_d^k(M)$ is called the {\it
$k$-th de~Rham cohomology space} of $M$.
The direct sum
$$H^*_d(M)=  \mathop{\oplus}\limits_{k=0}^n H_d^k(M)$$
is called the {\it de~Rham cohomology} of $M$.

\medskip

{\sf 2.3. Dolbeault complex.}
Consider a  complex manifold $M$ of complex dimension $n$
and the space $\Omega^{p,q}(M)$ of smooth $(p,q)$-forms on $M$.
The exterior derivative $d$ on complex manifolds is splitted into a
direct sum of two differential operators
$\p$ and $\bar\p$ such that
$d=\p +\bar\p\ ,\ d^2=\p^2=\bar\p^2=\p\bar\p +\bar\p\p =0\ .$
These operators act on the space $\Omega^{p,q}(M)$
as follows:
$$\p : \Omega^{p,q}(M)\to\Omega^{p+1,q}(M)\quad ,\quad
\bar\p : \Omega^{p,q}(M)\to \Omega^{p,q+1}(M)\ ,$$
$$d: \Omega^{p,q}(M)\to\Omega^{p+1,q}(M)\oplus \Omega^{p,q+1}(M)\ .$$

Consider a sequence of homomorphisms
$$
\Omega^{p,0}(M)\stackrel{\bar\p}{\to} \Omega^{p,1}(M)
\stackrel{\bar\p}{\to}...\stackrel{\bar\p}{\to}  \Omega^{p,n}(M)\to 0\ .
$$
Since $\bar\p^2 =0$, the direct sum
$$
\Omega^{p,*}(M)= \mathop{\oplus}\limits_{q=0}^n \Omega^{p,q}(M)
$$
of the spaces $\Omega^{p,q}(M)$ is a differential complex, called the
{\it Dolbeault complex}.
A $(p,q)$-form $\vp$ is called $\bar\p$-closed if $\bar\p\vp =0$.
In the case $q=0$ such $\bar\p$-closed forms are called {\it holomorphic}.

Denote by
$$
Z^{p,q}_{\bar\p}(M) := \Ker\bar\p\cap\Omega^{p,q}(M)=\{\o\in
\Omega^{p,q}(M) \ \mbox{such that}\ \bar\p\o =0\}
$$
the space of $\bar\p$-closed $(p,q)$-forms, i.e. {\it Dolbeault
$q$-cocycles} in the language of differential complexes, and by
$$
B^{p,q}_{\bar\p}(M) := \im\bar\p\cap\Omega^{p,q}(M)=\{\tau\in
\Omega^{p,q}(M) \ \mbox{such that}\ \tau =\bar\p\nu \ \mbox{for some}\
\nu\in\Omega^{p,q-1}(M)\}
$$
the space of $\bar\p$-exact $(p,q)$-forms, i.e. {\it Dolbeault
$q$-coboundaries}. We also denote by $\ce^p(M):=Z^{p,0}_{\bar\p}(M)$
the space of holomorphic $(p,0)$-forms.

The quotient space $H^{p,q}_{\bar\p}(M)=Z^{p,q}_{\bar\p}(M)/
B^{p,q}_{\bar\p}(M)$ is called the {\it $(p,q)$-th Dolbeault cohomology
space} of $M$.
The direct sum
$$H^{p,*}_{\bar\p}(M)=  \mathop{\oplus}\limits_{q=0}^n
H^{p,q}_{\bar\p}(M)$$
is called the {\it Dolbeault cohomology} of $M$.

\medskip

{\sf 2.4. \v{C}ech complex}. For any smooth $n$-manifold $M$ it is  possible
to choose an open covering $\ \U =\{U_\a\}_{\a\in I}$ such that
each nonempty finite intersection of the open sets $U_\a$ is
diffeomorphic to an open ball in $\R^n$ (or biholomorphic to a Stein
manifold for complex $n$-manifolds $M$)[12,13]. Such a covering will
be called a {\it good covering.} For a good covering the Poincar\'e
lemma holds on each finite intersection of the open sets $U_\a$, $\a\in I$.

 An ordered collection
$\langle U_{\a_0},...,U_{\a_m}\rangle$ of $m+1$ open sets from the
covering $\U$ such that $U_{\a_0}\cap ...\cap U_{\a_m}\ne\vn$ is
called an {\it $m$-simplex}. The set
$U_{\a_{0}...\a_m}:=U_{\a_0}\cap ...\cap U_{\a_m}$ is called a
{\it support} of the $m$-simplex $\langle U_{\a_0},...,U_{\a_m}\rangle$.

Let us consider the space $\cs$ of forms of a particular ``type"
defined {\it locally} on various open sets of a manifold $M$.
Depending on the structure of $M$ (smooth or complex-analytic)
this may be the space of smooth $k$-forms, holomorphic $(p,0)$-forms etc.
In other words, we consider various {\it sheaves} of forms over $M$ [12,13].

A {\it \v{C}ech $m$-cochain} $c^m$ with values in the space $\cs$ is a
collection $c^m=\{c_{\a_0...\a_m}\}$ of elements $c_{\a_0...\a_m}$ from
$\cs$ defined on supports $U_{\a_0...\a_m}$ of $m$-simplexes
$\langle U_{\a_0},...,U_{\a_m} \rangle$.
The space of \v{C}ech $m$-cochains for the covering $\U$ with
values in $\cs$ will be denoted by $C^m(\U , \cs )$.

Let us denote by $\rho_\a$ the restriction operator
acting on elements from $\cs$ as follows: if $f\in \cs$ is defined on
an open set $U$ then $\rho_\a f$ is defined on $U\cap U_\a$.
Now let us consider the map $\d :=\{\rho_{[\a ...]}\}$,
$$
\d : \{ c_{\a_0...\a_m}\}\to \{\rho_{[\a_0}c_{\a_1...\a_{m+1}]}\}\ ,
\eqno(5)
$$
where $c^m=\{c_{\a_0...\a_m}\}\in C^m(\U , \cs ), \ \d c^m=
\{\rho_{[\a_0 }c_{\a_1...\a_{m+1}]}\}\in
C^{m+1}(\U , \cs )$ and $[\a_0...\a_{m+1}]$ means antisymmetrization
w.r.t.~the indices $\a_0,...,\a_{m+1}$.
The operator $\d$ is called a {\it coboundary operator}.

Since $\rho_\a\rho_\b =\rho_\b\rho_\a$, we have $\d ^2=0$.
Therefore, one can consider a sequence of homomorphisms
$$
C^0(\U ,\cs )\stackrel{\d}{\to}...\stackrel{\d}{\to}C^{m-1}
(\U , \cs )\stackrel{\d}{\to}C^{m}(\U , \cs )
\stackrel{\d}{\to}C^{m+1}(\U , \cs )\stackrel{\d}{\to}...\ ,
$$
which gives the {\it \v{C}ech complex}
$
C^*(\U ,\cs )=\mathop{\oplus}\limits_{m\ge 0}C^m(\U ,\cs )\ .
$
The coboundary operator $\d$ is a ``differential" operator of this
complex in the terminology of differential complexes.

Denote by
$$
Z^m(\U , \cs ):= \Ker\d\cap C^m(\U ,\cs )=
\{ z\in C^m(\U , \cs ):\  \d z =0\}
\eqno(6)
$$
the {\it space of \v{C}ech $m$-cocycles},
and by
$$
B^m(\U , \cs ):= \im\d\cap C^m(\U ,\cs )=
\{ b\in C^m(\U , \cs ):\  b=\d c\ \mbox{for some}\ c\in C^{m-1}(\U ,\cs )\}
\eqno(7)
$$
the {\it space of \v{C}ech $m$-coboundaries}.

It is evident that
$
B^m(\U ,\cs )\su  Z^m(\U , \cs )  \su C^m(\U , \cs )\ .
$
Therefore, we can introduce the space
$H^m(\U ,\cs )= Z^m(\U , \cs ) /  B^m(\U ,\cs )$ of nontrivial \v{C}ech
$m$-cocycles,
where two distinct elements of $Z^m(\U ,\cs )$ are regarded as equivalent in
$H^m(\U ,\cs )$ if they differ by a coboundary.
We call $H^m(\U ,\cs )$ the {\it $m$-th \v{C}ech cohomology space} of
the covering $\U$ with coefficients in the space $\cs$.

The direct sum
$$
H^*(\U ,\cs )=\mathop{\oplus}\limits_{m\ge 0}H^m(\U ,\cs )
$$
is called the {\it \v{C}ech cohomology} of $\U$ with coefficients in $\cs$.
The result depends to some extent on the choice of covering $\U$, but
for a good covering this dependence disappear
and $H^m(\U ,\cs )\equiv H^m(M,\cs )$ (for proof see e.g. [12,13]).

\bigskip

{\bf 3. Field-theoretic description of de~Rham and Dolbeault
cohomologies}

\smallskip

{\sf 3.1. Abelian Chern-Simons and BF theories.}
Let $M$ be an oriented smooth $n$-manifold. The field-theoretic description
of the de~Rham  cohomology of $M$ can be given by the following
action functional [4]:
$$
S_{\rm dR}=\int_M\sum^\ell_{k=1}\o^{(n-k)}\wedge d\o^{(k-1)}\ ,
\eqno(8)
$$
where $\o^{(s)}\in \Omega^s(M)$ are $s$-forms on $M$, $s=0,1,...,n$,
and $\ell =\left [\frac{n+1}{2}\right ]$ is the integer part of the number
$\frac{n+1}{2}$.

The Euler-Lagrange equations for this action functional are
$$
d\o^{(k)}=0\ ,\quad k=0,1,...,n-1\ .
\eqno(9)
$$
Solutions to these equations are elements from $Z^k_d(M)$.
Exact forms from  $B^k_d(M)$ give trivial solutions.
Therefore, the space of nontrivial solutions (moduli space)
for the field equations (9) is given by the space
$\mathop{\oplus}\limits_{k=0}^{n-1}H^k_d(M)$.

The ``de~Rham theory"  described by $S_{\rm dR}$ generalizes
Abelian Chern-Simons theory defined by the action
$$
S_{\rm ACS}=\int_M A\wedge dA\ ,
\eqno(10)
$$
where $A:=\o^{(1)}\in\Omega^{1}(M)\ , \dim_\R M=3$.
One can compare $S_{\rm ACS}$ with $S_{\rm dR}$ for $n=3$:
$$
S_{\rm dR}=\int_M (\o^{(2)}\wedge d\o^{(0)} +
\o^{(1)}\wedge d\o^{(1)})\ .
$$
The ``de Rham theory" (8) generalizes also
Abelian topological BF theory with the action functional
$$
S_{\rm ABF}=\int_M B\wedge F\ ,
\eqno(11)
$$
where $B:=\o^{(n-2)}\in \Omega^{n-2}(M)\ ,\ F:=d\o^{(1)} ,\ 
\o^{(1)}\in\Omega^1(M)$.

\medskip

{\sf 3.2. Abelian holomorphic BF theory.}
Now we consider a complex $n$-manifold $M$. The field-theoretic
description of the Dolbeault cohomology of $M$ can be given by the following
action functional:
$$
S_{\rm Dol}= \int_M\sum_{q=1}^n\o^{(n-p,n-q)}
\wedge\bar\p\o^{(p,q-1)} \ ,
\eqno(12)
$$
where $\o^{(l,s)}\in\Omega^{l,s}(M)$. The equations of motion are
$$
\bar\p\o^{(p,q-1)}=0\ ,
\quad
\bar\p\o^{(n-p,n-q)}=0\ ,
\quad q=1,...,n\ .
\eqno(13)$$
The moduli space of solutions is a vector space
$
\mathop{\oplus}\limits_{q=1}^n\left (H^{p,q-1}_{\bar\p}(M)\oplus
H^{n-p,n-q}_{\bar\p}(M)\right )\ .
$
The ``Dolbeault theory" described by $S_{\rm Dol}$ generalizes
 Abelian holomorphic BF theory with the action functional
$$
S_{\rm AhBF}=\int_M B^{n,n-2}\wedge F^{0,2}\ ,
\eqno(14)
$$
where $B^{n,n-2}:= \o^{(n,n-2)} \in \Omega^{n,n-2}(M)$,
$F^{0,2}=\bar\p\o^{(0,1)}$, $\o^{(0,1)}\in \Omega^{0,1}(M)$.

\medskip

{\sf 3.3. Isomorphisms of cohomologies.} For a differentiable $n$-manifold,
we denote by $\Omega^k$ the space of locally defined $k$-forms
and by $\mr\subset \Omega^0$ the space of locally constant functions.
For a complex $n$-manifold,
by $\ce^p$ we denote the space of locally holomorphic $(p,0)$-forms
and by $\co =\ce^0$ the space of locally holomorphic functions.
In other words, we consider the sheaf $\mr$ of locally constant functions
and the sheaves $\Omega^k$, $\ce^p$ of forms.

\smallskip

{\sl Theorem 1.} The de~Rham cohomology $H^*_d(M)$ and \v{C}ech
cohomology  $H^*(M,\mr)$ of a differentiable $n$-manifold $M$
are isomorphic.

{}For proof see e.g. [12].

\smallskip

The de~Rham-\v{C}ech isomorphism reduces differential equations
$d\o^{(k)}=0$ on $k$-forms $\o^{(k)}$ (de Rham $k$-cochains) on
a smooth $n$-manifold $M$
to the functional equations $\rho_{[\a_0}c_{\a_1...\a_{k+1}]}=0$
on \v{C}ech $k$-cochains $\{c_{\a_0...\a_{k}}\}$ from $C^k(\U,\mr )$
defined on open subsets
$\{U_{\a_0...\a_k}\}$ of $M$. So, solutions of topological
CS and BF  theories can be described in terms of \v{C}ech cocycles.

\smallskip

{\sl Example 1.} Let us consider a de Rham 2-cocycle
$F\in Z^2_d(M)$, i.e. a $d$-closed 2-form $F$ on a
smooth $n$-manifold $M$: $dF=0$. For a   covering
$\U =\{U_\a\}_{\a\in I}$ of $M$, we have $F=\{\rho_\a F\}=
\{F_\a\}$ and $dF_\a =0$. By the Poincar\'e lemma [12],
each closed 2-form locally is exact and therefore on each open set
$U_\a$ there exists a 1-form $A_\a$ such that $F_\a =dA_\a$.

Let us introduce a collection $\{b_{\a\b}\}=\{\rho_\a A_\b - \rho_\b
A_\a\}$ of 1-forms defined on $U_\a\cap U_\b$.  It is easy to see
that they are  closed 1-forms:  $db_{\a\b}=\rho_\a F_\b - \rho_\b
F_\a =\rho_{[\a}\rho_{\b ]}F=0$.  Again, by the Poincar\'e lemma,
there exist {\it smooth} functions $s_{\a\b}$ such that
$b_{\a\b}=ds_{\a\b}$. Using a collection $\{s_{\a\b}\}$, we introduce
 functions $c_{\a\b\g}=\rho_{[\a}s_{\b\g ]}$ defined on
$U_\a\cap U_\b\cap U_\g\ne 0$.  One obtains
$dc_{\a\b\g}=\rho_{[\a}b_{\b\g ]}= \rho_{[\a}\rho_\b A_{\g ]}=0$,
i.e. $c_{\a\b\g}$ are {\it locally constant} functions. At the same
time we have $(\d c)_{\a\b\g\s}=\rho_{[\a}c_{\b\g\s ]}= \rho_{[\a}
\rho_\b s_{\g\s ]}=0$, i.e. $c=\{c_{\a\b\g}\}$ is a \v{C}ech
2-cocycle with values in the space $\mr$ of locally constant
functions.  Thus, to the de Rham 2-cocycle $F=\{F_\a\}\in Z^2_d(M)$
we have associated a \v{C}ech 2-cocycle  $c=\{c_{\a\b\g}\}\in
Z^2(\U ,\mr )$.

Conversely, for any  \v{C}ech 2-cocycle  $c=\{c_{\a\b\g}\}\in
Z^2(\U ,\mr )$ there exist collections $\{s_{\a\b}\}\in C^1(\U ,
\Omega^0)$ and $\{A_\a\}\in C^0(\U , \Omega^1)$ of smooth functions
and 1-forms such that
$$
c_{\a\b\g}=\rho_{[\a}s_{\b\g ]}\quad {\mbox {on}}\quad
U_\a\cap U_\b\cap U_\g\  ,
\qquad
ds_{\a\b}=\rho_{[\a}A_{\b ]}\quad {\mbox {on}}\quad
U_\a\cap U_\b \ .
$$
If we find local 1-forms $\{A_\a\}$ from these equations,
then a collection $\{dA_\a\}$  gives a $d$-closed 2-form
$F=\{dA_\a\}$ on $M$.

\smallskip

{\sl Theorem 2.} The Dolbeault cohomology  $H^{p,*}_{\bar\p}(M)$
and \v{C}ech cohomology $H^*(M,\ce^p)$
of a complex $n$-manifold $M$ are isomorphic.

{}For proof see e.g. [13].

\smallskip

The Dolbeault-\v{C}ech isomorphism reduces differential equations
(13) on $(p,q)$-forms (Dolbeault $q$-cochains) defined
on a complex $n$-manifold $M$ to functional equations (6) on \v{C}ech
$q$-cochains from $C^q(\U,\ce^p)$ defined on open subsets
$\{U_{\a_0...\a_q}\}$ of $M$. By using this isomorphism,
one obtains another description of solutions of holomorphic CS and
BF theories.

\smallskip

{\sl Example 2.}
Let us consider a Dolbeault 1-cocycle $A^{0,1}\in
Z^{0,1}_{\bar\p}(M)$, i.e a smooth
$\bar\p$-closed (0,1)-form $A^{0,1}$ on a complex $n$-manifold $M$:
$\bar\p A^{0,1}=0$. Let $\U =\{U_\a\}_{\a\in I}$ be a   covering of
$M$. Then $A^{0,1}$ can be represented by a collection
$\{A^{0,1}_\a\}$ of $\bar\p$-closed (0,1)-forms $A^{0,1}_\a :=\rho_\a
A^{0,1}$ on $U_\a$, $\a\in I$. By the Poincar\'e lemma [13], on each
open set $U_\a$ there exists a smooth function $b_\a$ such that
$A^{0,1}_\a=\bar\p b_\a$.

It is easy to see that on each
intersection  $U_\a\cap U_\b\ne\vn$ we have $\bar\p (\rho_\a
b_\b-\rho_\b b_\a)=\rho_\a A^{0,1}_\b-\rho_\b A^{0,1}_\a = (\rho_\a
\rho_\b-\rho_\b\rho_\a )A^{0,1}=0$ since $\rho_{[\a}\rho_{\b ]}=0$.
So, we obtain a collection $b=\{b_\a\}\in C^0(\U ,\Omega^{0})$ of
{\it smooth} functions $b_\a$ defined on $U_\a$ and a collection
$h=\{h_{\a\b}\}\in C^1(\U,\co )$ of {\it holomorphic} functions
$h_{\a\b}:=\rho_\a b_\b-\rho_\b b_\a$ defined on $U_\a\cap U_\b$, $\a
,\b\in I$.  It is not difficult to see that $h$ is a \v{C}ech
1-cocycle from $Z^1 (\U ,\co )$:  $(\d h)_{\a\b\g} = \rho_{[\a}
h_{\b\g ]}=\rho_{[\a}\rho_{[\b} b_{\g]]} =0$.  Thus, to the Dolbeault
1-cocycle $A^{0,1}$ we have associated a \v{C}ech 1-cocycle $h$.

Conversely, for any \v{C}ech 1-cocycle
$h=\{h_{\a\b}\}\in Z^1(\U,\co )$  there exists
a collection $b=\{b_\a\}\in C^0(\U , \Omega^{0})$ of smooth functions
$b_\a$ such that
$$
h_{\a\b}=\rho_\a b_\b -\rho_\b b_\a
\eqno(15)
$$
on $U_\a\cap U_\b$.  If we find such functions $\{b_\a\}$ then
a collection $\{\bar\p b_\a\}$ will give us a global $\bar\p$-closed
(0,1)-form $A^{0,1}=\{\bar\p b_\a\}$ on $M$. So, the isomorphism
between $H^{0,1}_{\bar\p}(M)$ and $H^1(M,\co )$ permits one to reduce
differential equations $\bar\p A^{0,1} =\{\bar\p A^{0,1}_\a\}=0$
to functional equations (15).

\bigskip

{\bf 4. Non-Abelian Chern-Simons and BF theories}

\smallskip

{\sf 4.1. Chern-Simons theories.}
Let $M$ be an oriented smooth 3-manifold, $\U =\{U_\a\}$ a good
covering of $M$, $G$ a matrix Lie group, and $\cg$ its Lie
algebra. Denote by $A$ a connection 1-form on a (topologically trivial)
principal $G$-bundle $P$ over $M$. For such bundles $A$ is a $\cg$-valued
1-form on $M$.

Consider the action functional of non-Abelian topological
Chern-Simons theory,
$$
S_{\rm CS} =\int_M \Tr (A\wedge dA + \frac{2}{3}A\wedge A\wedge A)\ ,
\eqno(16)
$$
and its field equations
$$
F_A\equiv dA + A\wedge A=0 \ ,
\eqno(17)
$$
called {\it zero curvature} equations. Solutions $A$ to eqs.(17)
define {\it flat} connections on $P$, i.e. differential operators
$d_A=d+A$ such that $d_A^2=0$.	 Analogously, non-Abelian holomorphic
Chern-Simons theories
are defined on complex 3-manifolds and describe flat (0,1)-connections
(holomorphic structures) [8-10].

Solutions $A$ of eqs.(17) can be considered
as de Rham 1-cocycles in the non-Abelian de Rham cohomology.
On each open set $U_\a$  eqs.(17) are solved trivially:
$A=\{A_\a\}$, $A_\a =\psi^{-1}_\a d \psi_\a$, where $\psi =\{\psi_\a\}$
is a collection of smooth $G$-valued functions on $\{U_\a\}$.
To obtain a global solution $A$ on $M$ from local solutions
$A_\a =\psi^{-1}_\a d \psi_\a$, one should solve the differential equations
$$
\psi^{-1}_{\a |\b} d \psi_{\a |\b} -\psi^{-1}_{\b |\a}
d \psi_{\b |\a} =0 \quad \mbox{on each intersection}\quad
U_\a\cap U_\b\ne\vn\ ,
\eqno(18)
$$
which simply mean that $\rho_{\b}A_\a = \rho_{\a}A_\b$ on
$U_\a\cap U_\b\ne\vn$. Here $\psi_{\a |\b}:=  \rho_{\b}\psi_\a$.
Equations (18) are equivalent to the equations
$$
d(\psi_{\a |\b}\psi^{-1}_{\b |\a} )=0\ .
\eqno(19)$$
We see that $c_{\a\b}:= \psi_{\a |\b}\psi^{-1}_{\b |\a} $ is
a (locally) constant $G$-valued function defined on  $U_\a\cap U_\b$.

The collection $\{c_{\a\b}\}$ of $G$-valued functions is a
\v{C}ech 1-cocycle in the non-Abelian \v{C}ech cohomology [14],
where the cocycle conditions are
$$ (\rho_\g c_{\a\b})(\rho_\a c_{\b\g}) (\rho_\b c_{\g\a})
=1\ \mbox{on}\ U_\a\cap U_\b\cap U_\g\ne\vn \ .
\eqno(20)$$
Therefore, solutions of eqs.(17) can be obtained by splitting
{\it locally constant} $G$-valued functions $c_{\a\b}$ satisfying eqs.(20)
into a product of two {\it smooth} $G$-valued functions $\psi_\a$ and
$\psi_\b^{-1}$ defined on $U_\a$ and $U_\b$,
respectively. Then, by virtue of the de~Rham-\v{C}ech correspondence,
a collection $\{\psi_\a^{-1}d \psi_\a\}=:A$ gives a global solution to
eqs.(17).

\medskip

{\sf 4.2. Topological BF theories.}
A generalization of (topological) Chern-Simons theories
to arbitrary dimensions
is given by (topological) BF theories [4,6,7].
The action functional for non-Abelian topological BF theory has the
following form:  $$ S_{\mathrm {BF}}=\int_M\Tr(B\wedge F_A),
\eqno(21) $$ where $M$ is an oriented smooth $n$-manifold, $F_A$ is
the curvature of a connection 1-form $A$ on a topologically trivial
principal $G$-bundle
$P$ over $M$, and $B$ is a $\cg$-valued $(n-2)$-form on $M$.
The variation of the action (21) w.r.t. $B$ gives eqs.(17), and the
variation of this action w.r.t. $A$ gives the equations
$$ d_AB=dB+ A\wedge B - B\wedge A=0 \ . \eqno(22) $$
Thus, topological BF theories describe flat connections $d_A$ on a bundle
$P$ over $M$ and $\cg$-valued $d_A$-closed $(n-2)$-forms $B$ on $M$.

Constructing solutions of the field equations of topological CS and BF
theories in terms of deformation theory of locally constant (flat)
bundles is discussed in [16]. Differential equations (17) are equivalent
to functional equations (20) which are solved by $c_{\a\b}=
\psi_{\a |\b}\psi^{-1}_{\b |\a}$ for some smooth $G$-valued functions
$\{\psi_\a\}$, and for a flat connection  $A=\{\psi^{-1}_\a d\psi_{\a}\}$
eqs.(22) can be easily reduced to equations  $d(\psi B\psi^{-1})=0$
from standard de Rham cohomology. Symmetries of CS and BF topological
theories can be described in terms of \v{C}ech 1-cocycles with values in
the sheaf of locally constant maps of the space $M$ into the Lie group $G$.
For more details see [16].

\medskip

{\sf 4.3. Holomorphic BF theories.}
Let $M$ be a complex $n$-dimensional manifold, $G$ a complex
matrix Lie group, $\cg$ its Lie algebra, $P$ a
topologically trivial principal
$G$-bundle over $M$, $A$ a connection 1-form on $P$, and $F_A=
dA+A\wedge A$ its curvature. Consider holomorphic BF theories [11]
with the action functional
$$
S_{\rm{hBF}}=\int_M \Tr(B^{n,n-2}\wedge  F^{0,2}_A),
\eqno(23)
$$
where $B^{n,n-2}$ is a $\cg$-valued $(n,n-2)$-form on $M$ and
$F^{0,2}_A$ is the (0,2)-component of the curvature tensor
$F_A=F^{2,0}_A+F^{1,1}_A+F^{0,2}_A$.
The field equations for the action (23) are
$$
\bar\p A^{0,1}+ A^{0,1}\wedge A^{0,1}=0\ , \quad
\bar\p B^{n,n-2}+ A^{0,1}\wedge B^{n,n-2}- B^{n,n-2}\wedge A^{0,1}=0\ ,
\eqno(24)
$$
where $A^{0,1}$ is the (0,1)-component of a connection 1-form $A=A^{1,0}+
A^{0,1}$ on $P$. If a representation of $G$ in the complex
vector space $\C^r$ is given,  we can associate with $P$ the
complex vector bundle $E=P\times_G\C^r$ and use vector bundles in the
description of BF theories.

It follows from eqs.(24) that models (23) describe flat (0,1)-connections
$\bar\p_A = \bar\p + A^{0,1}$ on $G$-bundles over
complex $n$-manifolds $M$
and $\bar\p_A$-closed $\cg$-valued $(n,n-2)$-forms $B^{n,n-2}$ on $M$.
A procedure of constructing solutions to eqs.(24) and mapping
solutions into one another (dressing transformations) are discussed
in [10,17]. This cohomological method of solving eqs.(24) is
based on the equivalence of \v{C}ech and Dolbeault descriptions
of holomorphic bundles and is a generalization to arbitrary dimensions
of the dressing approach (Riemann-Hilbert problems)
to solving integrable equations in two dimensions.

\bigskip

{\bf 5. Manin-Ward integrable systems}

\smallskip

{\sf 5.1. Double fibrations.}
Let us consider complex manifolds $X$ and $Z$ with a correspondence
between them defined by a double
fibration
$$
\begin{array}{c}
Y\\
^\eta\swarrow\quad \searrow^\rho \\
Z\qquad X
\end{array} \quad ,
\eqno(25)
$$
where $Y$ is a complex submanifold in the direct product
$Z\times X$, and projections $\eta$ and $\rho$ are surjective
holomorphic maps. We assume that the fibres
of $\eta$ are connected and simply connected complex
manifolds, and the fibres of $\rho$ are supposed to be compact.

The double fibration (25) appears  in the Kodaira relative
deformation theory of compact submanifolds of complex
manifolds [25]. Points of $X$ describe a family of complex
submanifolds $\eta (\rho^{-1}(x))$ of $Z$, and points of $Z$ describe
a family of submanifolds $\rho (\eta^{-1}(z))$ of $X$.
Note that Kodaira's double fibrations are used in
reparametrization-invariant geometric quantization of bosonic
string theory (see [26] and references therein).

\smallskip

{\sf 5.2. Integrable distributions.}
Under some mild topological conditions on the embeddings
$\eta (\rho^{-1}(x))\hookrightarrow Z$ the manifold $X$
comes equipped with a family of torsion-free affine connections
which are integrable on submanifolds  $\rho (\eta^{-1}(z))$ of $X$
[27,28]. We assume that these conditions are satisfied and
$X$ has such a connection.

The manifold $Y$ has two integrable transversal distributions:
the distribution $\cv^{1,0}_\eta$ of holomorphic vertical vector fields
in the fibration $\eta : Y\to Z$ and the distribution $\cv^{1,0}_\rho$
of holomorphic vertical vector fields in the fibration $\rho :
Y\to X$.

Let $\ce^1_\eta$ denotes the space of holomorphic
(1,0)-forms {\it dual} to the holomorphic vectors
from the distribution $\cv^{1,0}_\eta$. Since we have a
projection $\ce^1\to\ce^1_\eta$, we can introduce
a {\it relative} exterior derivative $\p_\eta$ as the
composition $\co\stackrel{\p}{\to} \ce^1 \to \ce^1_\eta$,
where $\ce^1$ and $\co$ are the sheaves of holomorphic
(1,0)-forms and holomorphic functions on $Y$, respectively.

\smallskip

{\sf 5.3. Integrable systems.}
Let us consider the double fibration (25), a topologically
trivial holomorphic rank $r$ vector bundle $E$ over $Z$,
the pulled-back bundle $\widetilde E\equiv \eta^*E$ over $Y$ and
the direct image bundle $\hat E\equiv\rho_*\eta^*E$ over $X$.
In local holomorphic trivializations, the bundle $\widetilde E$
is defined by holomorphic transition functions constant along
the fibres of $\eta$ (i.e. annihilated by $\bar\p$ and $\p_\eta$).
So, $\widetilde E$ is equipped with the holomorphic
structure $\bar\p$ and a flat relative
holomorphic (1,0)-connection $\p_\eta$: $\bar\p^2=0$, $\p_\eta^2=0$
 and $\p_\eta\bar\p + \bar\p\p_\eta=0$.

Suppose that the bundle $E$ is holomorphically trivial on
each submanifold $\eta (\rho^{-1}(x))$ of $Z$,
which is equivalent to saying that the bundle $\widetilde E=\eta^* E$
is holomorphically trivial on each submanifold $\rho^{-1}(x)$ of $Y$,
$x\in X$.
Such bundles  are called {\it X-trivial} holomorphic bundles [19].
Yu.Manin [19] and R.Ward [20] have shown that there is a one-to-one
correspondence between $X$-trivial holomorphic vector bundles $E$ on $Z$
and vector bundles $\hat E=\rho_*\eta^*E$
on $X$ with a differential operator $D: \hat E\to\rho_*\ce^1_\eta\otimes
\hat E$ flat on each submanifold $\rho (\eta^{-1}(z))$ of $X$, i.e.
$D^2|_{\rho (\eta^{-1}(z))}=0$ for all $z\in Z$. The operator
$D$ is constructed from  differential
operators on $X$ and the equations $D^2|_{\rho (\eta^{-1}(z))}=0$ yield
a system of nonlinear integrable differential
equations on $X$ [19-23].

\smallskip

{\sf 5.4. Dressing symmetries.}	Let us consider an $X$-trivial holomorphic
rank $r$ vector bundle $E\to Z$ and the pulled-back  bundle
$\widetilde E={\eta^*}E \to Y$ with the flat relative (1,0)-connection
$\p_\eta$. There is a one-to-one
correspondence between Manin-Ward integrable systems on $X$ and the
bundles $\widetilde E$. These bundles are locally constant (flat) along the
fibres of $\eta: Y \to Z$ and holomorphically trivial along the fibres
of $\rho: Y \to X$. Therefore, maps of solutions of the Manin-Ward
integrable equations into one another (dressing symmetries) can be
described by combining results on symmetries of flat [16] and holomorphic
[17] bundles. Namely, consider a covering $\U =\{U_\a\}$ of
$Y$ and holomorphic transition functions $\{f_{\a\b}\}$ in the
topologically trivial vector bundle $\widetilde E$ satisfying
$\p_\eta f_{\a\b}=0$. Let $\ch$ be the sheaf of
holomorphic $gl(r,\C)$-valued functions on $Y$ and $\ch_\eta $
its subsheaf of functions annihilated by $\p_\eta$. We
consider the Lie algebra $C^1(\U, \ch_\eta)$ of \v{C}ech
1-cochains with values in $\ch_\eta$ and define its action
on the transition functions $\{f_{\a\b}\}$ in $\widetilde E$ by
the formula
$$\d_\th f_{\a\b}= \th_{\a\b}f_{\a\b}-f_{\a\b}\th_{\b\a}\ ,$$
where $\{\th_{\a\b}\}\in C^1(\U , \ch_\eta)$. Following
[16,17] one can show that this action generates infinitesimal
deformations of the bundle $\widetilde E$ on $Y$ preserving the condition
of $X$-triviality and therefore descends to symmetries of Manin-Ward
integrable systems on $X$.
The algebra $C^1(\U , \ch_\eta )$ of collections $\{\th_{\a\b}\}$
with pointwise commutators generalizes affine Lie algebras
generated by algebra-valued holomorphic \v{C}ech 1-cochains
on the Riemann sphere $\C P^1$.

\smallskip

{\sf 5.5. Dolbeault description}. Manin-Ward integrable systems can be
described in terms of holomorphic BF theories on $Y$.
For this, let us consider smooth local trivializations of the bundle
$\widetilde E\to Y$ such that transition functions of
$\widetilde E$ become equal to unity on any intersection of charts.
This is possible since $\widetilde E$ is equivalent
to a product bundle. For a fixed covering $\U =\{U_\a\}_{\a\in I}$
of the manifold $Y$ and transition functions $f=\{f_{\a\b}\}$,
the change of trivialization
for $\widetilde E$ from holomorphic to smooth is described
by a collection $\psi =\{\psi_\a\}$ of smooth $GL(r,\C )$-valued
functions $\psi_\a$ on $U_\a$ such that $f_{\a\b}=\psi_\a\psi^{-1}_\b$.

After smooth mapping $\widetilde E$ onto the direct product bundle
$Y\times \C^r$, the holomorphic structure $\bar\p$ and the flat relative
connection $\p_\eta$  become
a holomorphic structure $\nabla^{0,1}=\bar\p +A^{0,1}=
\bar\p +\psi^{-1}\bar\p\psi$ and a flat relative (1,0)-connection
$\nabla^{1,0}_\eta =\p_\eta + A^{1,0}_\eta =\p_\eta +\psi^{-1}\p_\eta\psi$
such that
$$
(\nabla^{0,1})^2=\bar\p A^{0,1} + A^{0,1}\wedge A^{0,1}=0\ ,\quad
\p_\eta A^{0,1} + \bar\p A^{1,0}_\eta + A^{0,1} \wedge A^{1,0}_\eta +
A^{1,0}_\eta \wedge A^{0,1}=0\ ,
\eqno(26a)
$$
$$
(\nabla^{1,0}_\eta )^2=\p_\eta A^{1,0}_\eta + A^{1,0}_\eta \wedge
A^{1,0}_\eta =0\ .
\eqno (26b)
$$
Equations (26a,b) mean that the bundle $\widetilde E\simeq Y\times\C^r$
is holomorphic and its restrictions to the fibres $\eta^{-1}(z)$ of
the projection $\eta : Y\to Z$ are locally constant (flat) bundles
for any $z\in Z$. Functions $\psi_\a\in GL(r,\C )$ defining these bundles
can be chosen to be holomorphic in complex coordinates on $U_\a\cap
\eta^{-1}(z)$ for any $z\in Z$, and therefore we obtain $A^{0,1}_\eta =0$.

Recall that the bundle $\widetilde E\to Y$ is $X$-trivial.
Up to a gauge transformation this is equivalent to the equations
$$A^{0,1}_\rho =0\ ,
\eqno(26c)$$
where $A^{0,1}_\rho $ is the component of $A^{0,1}$ along the distribution
$\cv^{0,1}_\rho$ of antiholomorphic vertical vector fields in the
fibration $\rho : Y\to X$.
In other words, $GL(r,\C )$-valued functions $\psi_\a$ from a
collection $\psi =\{\psi_\a\}$ and $A^{1,0}_\eta =\psi^{-1}\p_\eta\psi$
can be chosen to be holomorphic in complex coordinates on
$U_\a\cap\rho^{-1}(x), x\in X$. Equations (26) give the
Dolbeault description of $X$-trivial holomorphic bundles $\widetilde E=
\eta^*E\to Y$.

A restriction of the flat relative (1,0)-connection
$\nabla^{1,0}_\eta$ to $\rho^{-1}(x)\hookrightarrow Y$ is a differential
operator with values in a holomorphic vector bundle over $\rho^{-1}(x)$.
It can be expanded in global holomorphic sections of this bundle
 and projected to $X$. Then we obtain an operator
$D\equiv\rho_*\nabla^{1,0}_\eta$ defining Manin-Ward integrable equations
on $X$. These equations arise as the compatibility conditions
$(\nabla^{1,0}_\eta )^2=0$ (equivalent to $D^2|_{\rho (\eta^{-1}(z))}=0$)
for the linear system of equations
$$\nabla^{1,0}_\eta \vp =0\ ,
\eqno(27)$$
where $\vp$ is a section of the bundle $\widetilde E$ on $Y$, depending on
$x\in X$ and $\lambda\in \rho^{-1}(x)$. Parameters $\l$
are called {\it spectral parameters}. For more details see [19-24].

\smallskip

{\sf 5.6. Ward's example} [20].
Here $X$ is an open subset of the complex vector
space $\C^{k(m+1)}$, $Y=X\times \C P^m$ and
$Z$ is an open subset of the complex projective space $\C P^{m(k+1)}$ (see
the diagram (25)). The space $Y$ is a $\C P^m$-bundle over $X$ with the
canonical holomorphic projection $\rho : Y\to X$. Let $\lambda^j$
($j=0,...,m$) be homogeneous coordinates on $\C P^m$ and $x^{aj}$
($a=1,...,k$) complex coordinates on $X\subset \C^{k(m+1)}$. Then
the integrable
distribution $\cv^{1,0}_\eta$ on $Y$ is spanned by the vector fields
$V_a =\lambda^j \p_{aj}$, $\p_{aj}:={\p}/{\p x^{aj}}$.
The space $Z$ is the quotient of $Y$ by the flows of the
vector fields $V_a$, $\dim_\C Z=m(k+1)$. For each fixed
$\lambda=\{\lambda^j\}\in \C P^m$, the vectors $V_1(\lambda ), ...,
V_k(\lambda )$ span a complex $k$-plane $\a (\lambda)$ in $X$.

There is a one-to-one correspondence between $X$-trivial holomorphic
vector bundles $E$ over $Z$ and bundles $\hat E = \rho_*\eta^*E$
over $X$ with a connection $D= dx^{aj}D_{aj}= dx^{aj}(\p_{aj}+ A_{aj})$
flat on each complex $k$-plane $\a (\l )$, $\l\in \C P^m$.
Here $A_{aj}(x)$ are matrix-valued functions on $X$ (components of a
connection 1-form on $\hat E$). The operator $\nabla^{1,0}_\eta$
is defined by the differential operators
$\nabla^{1,0}_a := \l^j D_{aj}$ and the
linear system of equations is $\nabla^{1,0}_a \vp$=0,
where $\vp (x,\l )$ is a section of the bundle $\eta^*E \to Y$.
The compatibility conditions of this linear system are polynomials in
$\lambda^j$ :
$$\lambda^j\lambda^l [\p_{aj}+ A_{aj},\p_{bl}+A_{bl}] =0\ .$$
After equating each
coefficient of these polynomials to zero we obtain a set of nonlinear
differential equations on the matrix-valued functions $A_{aj}(x)$ on $X$.

\smallskip

{\sf 5.7. Supersymmetric Yang-Mills}.
It is not difficult to introduce supersymmetric generalizations of
Manin-Ward integrable systems [19,21]. One of the most interesting
examples of such integrable models is the supersymmetric Yang-Mills
theory in ten dimensions [29]. For supersymmetric Yang-Mills model
on complexified $d=10$ space-time, the space $X$ is a region in the
superspace $\C^{10|16}$ (10 bosonic and 16 fermionic coordinates),
$Y$ is a complex supermanifold of dimension $18|16$ and $Z$ is a
complex supermanifold of dimension $17|8$ (see [21,29,30,31] for
more details and references).  Importance of this model is
connected with the fact that it appears in the low-energy limit of
superstrings. The \v{C}ech approach and advocated
cohomological methods may be helpful in describing moduli spaces and
symmetries of $d=10$ supersymmetric Yang-Mills and superstring theories.

\bigskip

{\bf Acknowledgements}

\smallskip

This work is supported in part by the grant RFBR-99-01-01076
and the Heisenberg-Landau Program. T.A.I. is grateful to the
Institut f\"ur Theoretische Physik der Universit\"at Hannover
for hospitality and support during the final stage of the work.

\vspace{5mm}

{\bf References}

\begin{list}{}{\setlength{\topsep}{0mm}\setlength{\itemsep}{0mm}%
\setlength{\parsep}{0mm}}
\item[1.]
E.Witten,
Commun.Math.Phys. {\bf 117} (1988) 353.

\item[2.]
D.Birmingham, M.Blau, M.Rakowski and G.Thompson,
Phys.Rep. {\bf 209} (1991) 129;
B.Dubrovin, Lect.Notes in Math. {\bf 1620} (1996) 120.

\item[3.]
J.M.F.Labastida, Chern-Simons gauge theory: ten years after,
hep-th/9905057;
A.S.Cattaneo and C.A.Rossi, Higher-dimensional BF
theories in the \\Batalin-Vilkovisky formalism: the
BV action and generalized Wilson loops,\\ math.qa/0010172.

\item[4.]
A.S.Schwarz,
Lett.Math.Phys. {\bf 2} (1978) 247;
Commun.Math.Phys. {\bf 67} (1979) 1.

\item[5.]
E.Witten,
Commun.Math.Phys. {\bf 121} (1989) 351.

\item[6.]
G.T.Horowitz,
Commun.Math.Phys. {\bf 125} (1989) 417.

\item[7.]
M.Blau and G.Thompson,
Ann.Phys. {\bf 205} (1991) 130.

\item[8.]
  E.Witten,
%Chern-Simons gauge theory as a string theory,
In: The Floer memorial volume, Progr. Math. 133 (Birkh\"{a}user, Boston,
1995) p.637  [hep-th/9207094].

\item[9.]
M.Bershadsky, S.Cecotti, H.Ooguri and C.Vafa,
Commun.Math.Phys. {\bf 165} (1994) 311 [hep-th/9309140].

\item[10.]
A.D.Popov,
Nucl.Phys. {\bf B550} (1999) 585  [hep-th/9806239].

\item[11.]
A.D.Popov,
Phys.Lett. {\bf B473} (2000) 65  [hep-th/9909135].

\item[12.]
R.Bott and L.W.Tu, Differential forms in algebraic topology
(Springer, Berlin, 1982).

\item[13.]
P.Griffiths and J.Harris,
Principles of algebraic geometry (Wiley, New York, 1978).

\item[14.]
J.Frenkel, Bull.Soc.Math.France {\bf 85} (1957) 135;
P.Dedecker, Canad.J.Math. {\bf 12} (1960) 231;
A.L.Oni\v{s}\v{c}ik, Trudi Mosk.Mat.Obsch. {\bf 17} (1967) 45.

\item[15.]
A.Levin and M.Olshanetsky, Commun.Math.Phys. {\bf 188} (1997) 449\\
{[{alg-geom/9605005}]}.

\item[16.]
T.A.Ivanova and A.D.Popov,
J.Nonlin.Math.Phys. {\bf 7} (2000) 480 [hep-th/9908177].

\item[17.]
T.A.Ivanova and A.D.Popov,
J.Math.Phys. {\bf 41} (2000) 2604 [hep-th/0002120].

\item[18.]
A.D.Popov,
Rev.Math.Phys. {\bf 11} (1999) 1091 [hep-th/9803183];
T.A.Ivanova,
In: Moduli spaces in mathematics and physics (Hindawi
Publ. Corporation, 1999) p.77  [math-ph/9902015].

\item[19.]
Yu.I.Manin, Gauge field theory and complex geometry
(Springer, Berlin, 1988)\\ {[Russian: Nauka, Moscow, 1984]}.

\item[20.]
R.S.Ward, Nucl.Phys. {\bf B236} (1984) 381.

\item[21.]
M.M.Kapranov and Yu.I.Manin, Russian Math.Surveys, {\bf 41:5}
(1986) 85.

\item[22.]
R.J.Baston and M.E.Eastwood,
The Penrose transform: Its interaction with representation theory
(Clarendon Press, Oxford, 1989).

\item[23.]
 L.J.Mason  and N.M.J.Woodhouse,
 Integrability, self-duality, and twistor theory
(Clarendon Press, Oxford, 1996).

\item[24.]
R.S.Ward, Phil.Trans.R.Soc.Lond. {\bf A315} (1985) 451,
Lect.Notes in Phys. {\bf 280} (1987) 106;
L.J.Mason and G.A.J.Sparling, Phys.Lett. {\bf A137} (1989) 29;
A.L.Carey, K.C.Hannabuss, L.J.Mason and M.A.Singer,
Commun.Math.Phys. {\bf 154} (1993) 25;
T.A.Ivanova and A.D.Popov,
 Theor.Math.Phys. {\bf 102} (1995) 280,
Phys.Lett. {\bf A205} (1995) 158 [hep-th/9508129].

\item[25.]
K.Kodaira, Ann.Math. {\bf 75} (1962) 146.

\item[26.]
A.D.Popov, Theor.Math.Phys. {\bf 83} (1990) 608;
Sov.J.Nucl.Phys. {\bf 53} (1991) 876; A.D.Popov and A.G.Sergeev,
In: Algebraic geometry and its applications (Vieweg, Wiesbaden, 1994)
p.136.

\item[27.]
S.A.Merkulov, Proc.Amer.Math.Soc. {\bf 124} (1996) 1499;\\
S.Merkulov and L.Schwach\-h\"ofer, Ann.Math. {\bf 150} (1999) 77.

\item[28.]
R.Bryant, Asterisque {\bf 266} (2000) 351 [math.dg/9910059].

\item[29.]
E.Witten,
 Nucl.Phys. {\bf B266} (1986) 245.

\item[30.]
E.Abdalla, M.Forger and M.Jacques,
Nucl.Phys.  {\bf B307} (1988) 198.

\item[31.]
J.-L.Gervais and M.V.Saveliev, Nucl.Phys. {\bf B554} (1999) 183
[hep-th/9811108].

\end{list}
\end{document}